\documentclass{pkas}
\def\year{2015} % publication year
\def\month{September} % publication month
\def\volume{30} % journal volume
\def\issue{2} % journal issue
\def\beginpage{145} % first page of article
\def\endpage{148} % last page of article
\def\received{October 31, 2014} % date paper was received by PKAS
\def\accepted{July 31, 2015} % date of acceptance
\date{Received \received ; accepted \accepted}

\def\simlt{\lower.5ex\hbox{$\; \buildrel < \over \sim \;$}}
\def\simgt{\lower.5ex\hbox{$\; \buildrel > \over \sim \;$}}
\def\kms{km s$^{-1}$}

\def\msol{{$M_\odot$}}

\title{
Near-Infrared Spectroscopy of Young Galactic Supernova Remnants
}

\author{Bon-Chul Koo, Yong-Hyun Lee}
\affil{Seoul National University, Seoul 151-747, Korea; \email{koo@astro.snu.ac.kr}}
\def\corrauthor{
B.-C. Koo
}

\def\runningauthor{
B.-C. Koo
}

\def\runningtitle{
Near-Infrared Spectroscopy of Young Galactic Supernova Remnants
}

\def\keywords{
infrared: ISM: shock waves: supernova remnants
}

\def\abstracttext{
Young Galactic supernova remnants (SNRs) are where we can observe closely
the supernova (SN) ejecta and its interaction with circumstellar/interstellar
medium. Therefore, they provide an opportunity to explore the explosion
and the final stage of the evolution of massive stars.
Near-infrared (NIR) emission lines in SNRs mostly originate from
shocked dense material. In shocked SN ejecta, forbidden lines from
heavy ions are prominent, while in shocked circumstellar/interstellar 
medium, [Fe II] and H$_2$ lines are prominent. [Fe II] lines are strong 
in both media, and therefore [Fe II] line images provide a good starting 
point for the NIR study of SNRs. There are about twenty SNRs detected 
in [Fe II] lines, some of which have been studied in NIR spectroscopy.   
We will review the NIR [Fe II] observations of SNRs and introduce  
our recent NIR spectroscopic study of the young core-collapse 
SNR Cas A where we detected strong [P II] lines. 
}

\begin{document}
\pkashead %% set title, authors, abstract, etc.

%%%%%%%%%%%%%%%%%%%%%%%%%%%%%%%%%%%%%%%%%%%%%%%%%%%%%%%%%%%%%%%%%%%%%
%%% BEGIN MAIN TEXT HERE %%%%%%%%%%%%%%%%%%%%%%%%%%%%%%%%%%%%%%%%%%%%
%%%%%%%%%%%%%%%%%%%%%%%%%%%%%%%%%%%%%%%%%%%%%%%%%%%%%%%%%%%%%%%%%%%%%

\section{Introduction}

In 1980s, NIR spectra of several SNRs were obtained
\citep{seward1983, graham1987, burton1988, oliva1989, graham1990, oliva1990}. 
A surprising result was that the forbidden lines from Fe II are much 
stronger than hydrogen recombination lines in contrast to HII regions.  
Also, it was found that NIR ro-vibrational H$_2$ lines 
are bright in some SNRs, which might be interacting with 
molecular clouds \citep{burton1989, burton1993}. 
These early studies showed 
that [Fe II] and H$_2$ lines are sensitive probes 
of shocked gas in SNRs.

%These are one of the first spectra and H2 images, and show the difficulty 
%of NIR observations in early days.

More recently, it was found that the NIR spectra of 
young SNRs are interesting, 
sometimes showing many 
ionic lines from supernova (SN) ejecta material
\citep{gerardy2001, koo2007, moon2009, lee2013, koo2013}. 
In the Milky Way, there are about 300 SNRs, 
and 20--30 of them are younger than several thousand years. 
These young SNRs still have the imprints 
of explosion. So NIR spectroscopy can probe the explosion dynamics, 
SN nucleosynthesis, and the late-stage evolution of progenitor star.
Among young SNRs, G11.2$-$0.3, RCW 103, W49B, and Cas A are known to be 
bright in [Fe II] lines, and
they are all small core-collapse SNRs that might be interacting with dense CSM
\citep{koo2014}. 
We have recently completed a survey of the Galactic plane in [Fe II] 
1.644 $\mu$m emission line, called UWIFE 
(UKIRT Widefield Infrared survey for Fe$^+$) survey \citep[][see also the paper 
by Lee, J.-J. in this volume]{leejj2014}, where 
we detected [Fe II] emission in 17 SNRs out of 80 SNRs in the survey area 
($\ell=7^\circ$ to $65^\circ$ and $|b|\le 1.3^\circ$). Therefore, the SNR samples
with [Fe II] emission will increase in near future. 

In this paper, we will first briefly summarize the characteristics and applications of 
NIR [Fe II] lines and then introduce our recent NIR spectroscopic 
study of Cas A where we detected strong [P II] lines.

\section{NIR [Fe II] emission from SNRs}

[Fe II] line images of young SNRs show that [Fe II] line emitting regions are
filamentary and spatially confined in contrast to radio or X-rays
\citep[see Fig. 4 of][]{koo2014}. 
This is because they originate from different environments.
Figure 1 is a schematic diagram showing the structure of young SNRs. 
The SNR is bounded by 
SN blast wave or forward shock. 
The shocked ambient medium is confined between SN blast wave and contact discontinuity. 
In the innermost region, 
there is freely expanding SN ejecta, which is heated when it encounters the reverse shock. 
And radio and X-rays are emitted from these shocked SN ejecta and shocked ambient medium, 
which is mostly circumstellar medium (CSM) 
in the case of young SNRs. In contrast, the NIR emission is emitted when 
the shocks encounter either dense ambient medium or dense SN ejecta and becomes 
radiative. 

In shocks propagating into a medium of normal abundance, an 
extended temperature plateau region at $T_e\sim 10^4$~K 
develops in postshock cooling region, and 
this is where [Fe II] lines are emitted \citep{koo2014}. 
In shocks propagating into SN ejecta, the temperature and 
emissivity profiles are very different 
because of strong cooling \citep{koo2013}. 
Fe II has 16 levels near the ground state 
that can be easily excited in the cooling region, and can emit strong NIR lines. 
In the NIR band, two strongest lines are 1.257 and 1.644 $\mu$m lines, 
and 10--20 lines are seen \citep[][and references therein]{koo2014}.

\begin{figure}
\centering
\includegraphics[width=80mm]{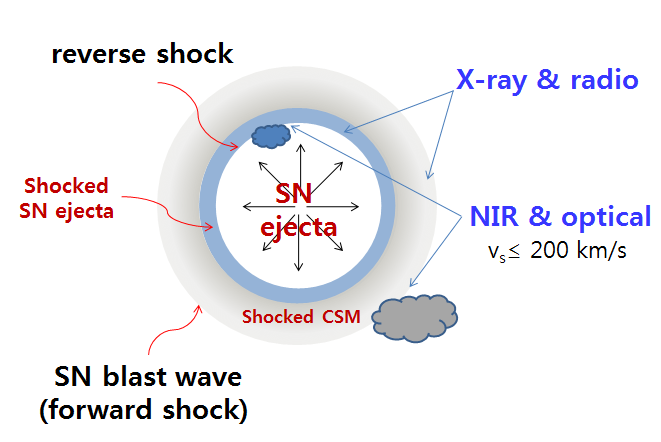}
\caption{Schematic diagram showing the structure of young SNRs and 
where emissions of different wavebands originate. \label{fig:pkasfig1}}
\end{figure}

\begin{figure}
\centering
\includegraphics[width=80mm]{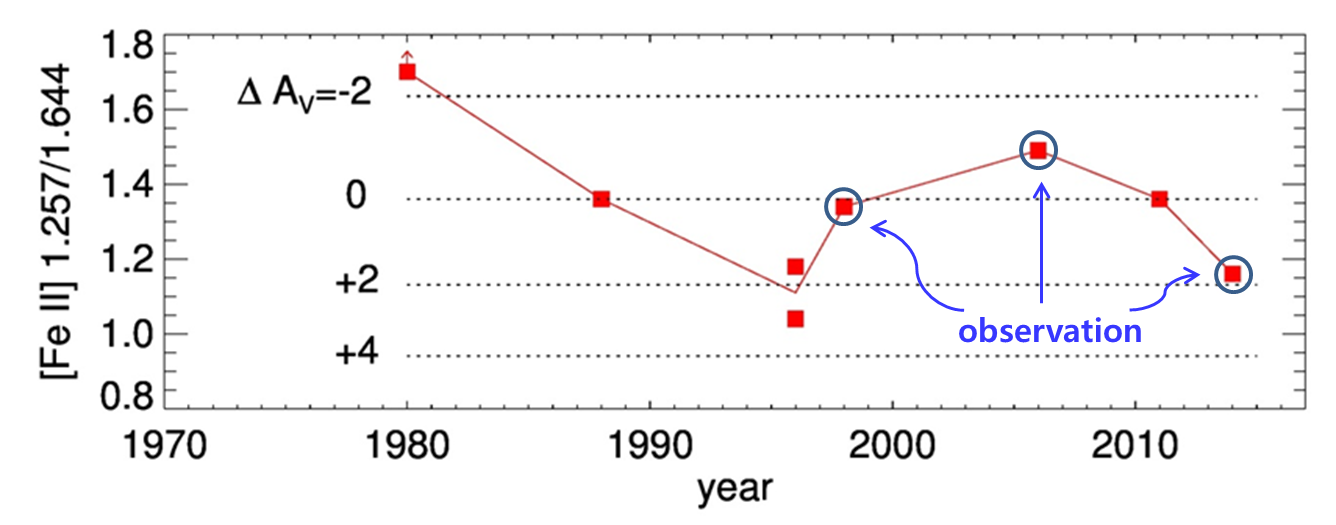}
\caption{
[Fe II] 1.257/[Fe II] 1.644 line intensity ratios reported in literature. The dotted lines mark 
the differences in $A_V$  when the [Fe II] 1.257/1.644 ratio different from 
1.36 is adopted. 
Referenecs: Nussbaumer \& Storey (1980), Nussbaumer \& Storey (1988),  
Quinet et al. (1996) (SST/HFR),  
Bautista \& Pradhan (1998) (observed, based on Everett \& Bautista 1996 private comm.),  
Smith \& Hartigan (2006), Deb \& Hibbert (2011), Giannini et al. (2014) \label{fig:pkasfig2}}
\end{figure}

\begin{figure}
\centering
% \vspace*{0.3in}
% \hspace*{-0.33in}
\includegraphics[width=80mm]{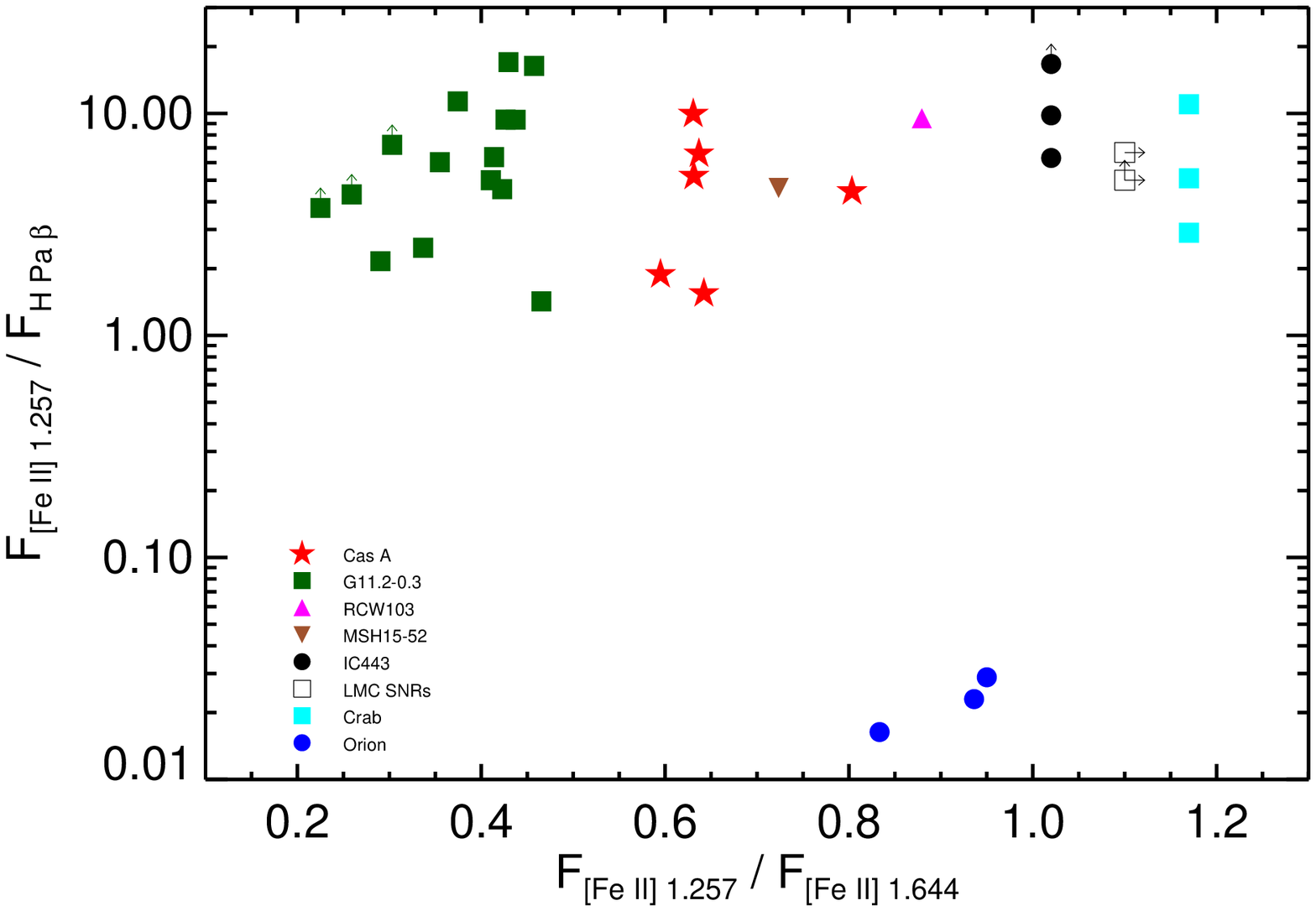}
\caption{
[Fe II] 1.257/H Pa-$\beta$ versus 
[Fe II] 1.257/[Fe II] 1.644 line intensity ratio of SNRs. 
Note that [Fe II] 1.257/[Fe II] 1.644 ratio is an indicator 
of extinction to the source. The ratios of Orion region are marked for comparison. 
%The scale bar on the right marks 
%theoretical [Fe II] 1.257/H Pa-$\beta$ intensity ratios from the Raymond 
%and the MAPPINGS shock codes for shocks of different velocities propagating into 
%a preshock density of $10^2$ cm$^{-3}$ \citep{koo2014b,allen2008}. 
Referenecs: Cas A \citep{gerardy2001, koo2013, lee2014}, 
G11.2-0.3 \citep{koo2007, moon2009}, RCW 103 \citep{oliva1989, oliva1990},
MSH 15-52 \citep{seward1983}, IC 443 \citep{graham1987}, 
LMC SNRs \citep{oliva2001}, Crab \citep{graham1990}, 
Orion \citep{walmsley2000}.
\label{fig:pkasfig3}}
\end{figure}

There are two direct applications of [Fe II] lines: 
(1) The extinction to the source can be accurately measured from the lines originating 
from the same 
upper levels, and (2) electron density of the emitting region can be derived 
using lines with comparable 
excitation energies.  A major uncertainty in these applications is from atomic constants. 
For example, Figure 2 shows 
[Fe II] 1.257/[Fe II] 1.644 line intensity ratios, which is a good 
indicator of extinction,  
predicted by theoretical calculations and also the ratios derived from observations. 
There is still some scatter and, depending on the adopted atomic constants, 
$A_V$ could be off by a few magnitudes.

As pointed out earlier, the ratio of [Fe II] lines to  H-recombination lines 
are much higher than those of HII regions, 
so that the ratio can be used to discriminate the 
shock-ionized gas from photoionized gas. 
Figure 3 shows that 
[Fe II] 1.257/H Pa-$\beta$ line intensity ratio of SNRs is greater than 0.1 while it 
is 0.01--0.03 in the Orion HII region.
The high ratio is partly 
due to the extended line-emitting 
region in the shock and probably also partly due to the increase of the gas-phase Fe abundance 
caused by the destruction of dust grains.
For SNRs, one can derive the shock speed by 
comparing the ratio with shock models. 
But the predicted ratios could be different among models, sometimes by a large factor,
so that one should be careful in doing such analysis.

\section{NIR [Fe II] and [P II] Emission from Cas A}

\subsection{NIR spectroscopy of Cas A}

Cas A is a 330-yr old young SNR. 
It is one of the few SNRs whose SN types are confirmed 
from the spectrum of light echo. 
It is SN IIb with a progenitor mass between 15 and 25 \msol. 
Recently, IR space missions detected a significant amount of dust in this remnant, 
which amplified the interests on this object.

From optical studies, it has been well known that there are two types of dense knots, 
namely FMKs which are SN ejecta knots moving at several 
thousand \kms\ and QSFs which are dense 
CS knots moving at a few hundred \kms. 
A few people carried out NIR spectroscopy to study the extinction to the remnant and also 
the density of knots \citep{gerardy2001, eriksen2009}. 
In particular, Gerardy \& Fesen showed that NIR spectral features of 
FMKs and QSFs are quite different.
We also obtained JHK long-slit spectra along the bright shell and 
identified 63 clumps \citep{koo2013, lee2014}. 
NIR spectra of the knots could be classified into three distinct groups: 
He-rich knots with strong He lines, S-rich knots with strong S 
and other intermediate mass element lines, 
and Fe-rich knots where essentially only Fe lines appear.
These spectroscopically-distinct knots also demonstrate 
different kinematic properties. He-rich knots are moving slowly, 
while S-rich and Fe-rich knots are moving at velocities as high as 2000 \kms. 
These spectroscopic and kinematic properties imply that He-rich knots are CS knots 
corresponding to QSFs while S-rich knots are SN ejecta knots corresponding to FMKs. 
Fe-rich knots are most likely SN ejecta material from the innermost core.

\subsection{[P II] lines in Cas A and SN Nucleosynthesis}

Among the NIR lines, an interesting line was 1.189 $\mu$m line from 
Phosphorus (P). It was detected mainly in S-rich knots and the line intensities 
were as strong as those of [Fe II] lines.
This is surprising because P is not an abundant element. 
Its cosmic abundance is $X({\rm P/H})=2.8\times 10^{-7}$ 
and its relative abundance to Fe by number is only 1/110. 
[P II] 1.189 $\mu$m line 
originates from a level whose excitation energy is almost identical to that of 
[Fe II] 1.257 $\mu$m line, and therefore the two lines originate 
from almost the same region in shocked gas. And one can derive the relative abundance of
the two elements directly from the line intensity ratios without a detailed 
analysis using a shock code. 
For the range of electron densities of Cas A knots, i.e., 
$n_e=3\times 10^3$ to $2\times 10^5$~cm$^{-3}$,  
$F_{\rm [P II]1.189}/F_{\rm [Fe II]1.257}= (3-7) X({\rm P/Fe})$ 
\citep[][see also Oliva et al. 2001]{koo2013}.
The derived $X({\rm P/H})$ of He-rich knots were close to the solar abundance 
while the abundance of S-rich knots were as much as 100 times higher than 
the cosmic abundance. This confirmed the in situ production of P. 
And the observed abundance ratios nicely fit into the theoretical range of 
15 \msol\ SN. However, the ratios were higher that the ratio for a complete mixing of core 
below He-rich layer, which implied that these dense knots largely retained  
their original abundances. This was the first result confirming the 
nucleosynthesis of P in SN.

\subsection{[P II] lines for the ISM Study}

[P II] 1.189/[Fe II] 1.257 intensity ratio can be used for the study of shock processing of dust 
in the ISM because P is not depleted while Fe is mostly locked in dust grains. 
In the general ISM, therefore, 
[P II] 1.189/[Fe II] 1.257 intensity ratio is high. But in fast shocks, 
dust grains are destroyed,  
so that the gas phase Fe abundance increases and 
therefore [P II]/[Fe II] ratio decreases. 
These properties have been used for the study of dust processing in HH objects 
and the origin of [Fe II] emission in external galaxies 
\citep[e.g.,][]{garcia2010, oliva2001}. 
Figure 4 shows ${F_{\rm [P II]1.189}/F_{\rm [Fe II]1.644}}$
versus ${F_{\rm [FeII]1.257}/F_{{\rm H Pa}-\beta}}$ of 
SNRs, HII regions, HH objects, and the galaxy NGC 1068. Note that both ratios are shock indicators, 
so that SNRs and HII regions are well separated in this diagram. NGC 1068, 
however, has [P II] 1.189/[Fe II] 1.257 intensity ratio close to HII region, but 
[Fe II] 1.257/H Pa-$\beta$ intensity ratio between those of HII region and SNRs.

\section{Summary}

The following is a brief summary:

\noindent
(1) NIR spectra of SNRs are dominated by [Fe II] and H$_2$ emission lines 
unless the emission is from SN ejecta material. 
[Fe II] lines can be used to derive $A_V$, $n_e$, and shock parameters. 
But there is an uncertainty associated with atomic parameters and shock codes.
[Fe II] lines can be used to infer the environment of SNRs, e.g., 
the SNRs which are interacting with dense CSM are bright in [Fe II] lines.

\noindent
(2) NIR spectroscopy of Cas A shows that CS and SN ejecta knots are clearly 
distinguished in their NIR spectra. [P II] lines are strong in S-rich SN ejecta 
material, and the derived X[P/Fe] is compatible with the local nucleosynthetic SN model.
[P II] line provides an accurate X(P/Fe), which can be used to study the shock 
destruction of dust grains and the origin of [Fe II] emission lines.

%%% ACKNOWLEDGMENTS (IF ANY) %%%%%%%%%%%%%%%%%%%%%%%%%%%%%%%%%%%%%%%%

\acknowledgments
This research was supported by Basic Science Research Program through the National 
Research Foundation of Korea(NRF) funded by the Ministry of Science, ICT and 
future Planning (2014R1A2A2A01002811). 

%%% FIGURE %%%%%%%%%%%%%%%%%%%%%%%%%%%%%%%%%%%%%%%%%%%%%%%%%%%%%%%%%%%%%%%%%%%%
\begin{figure}
\centering
\includegraphics[width=80mm]{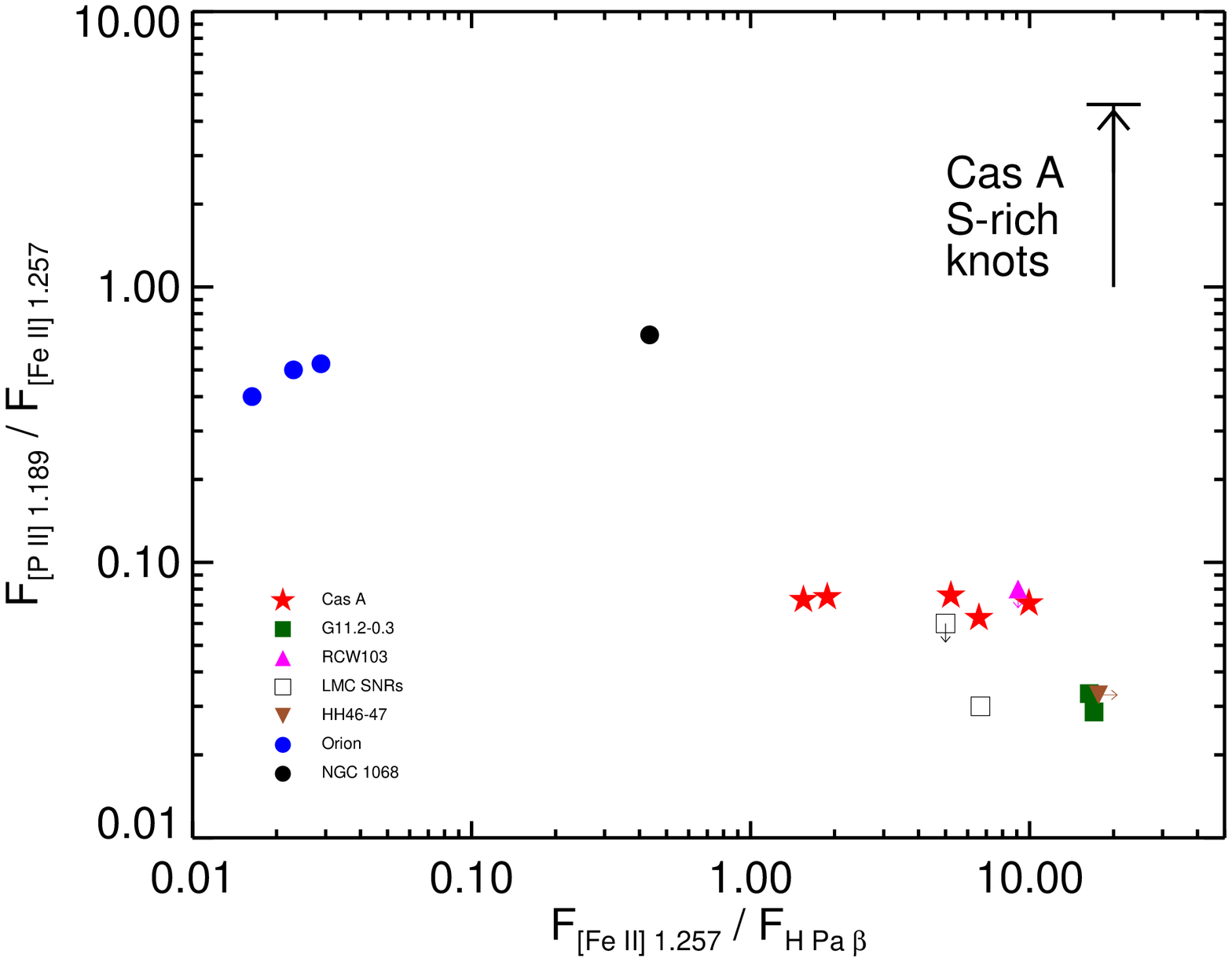}
\caption{
[P II] 1.189/[Fe II] 1.257 versus [Fe II] 1.257/H Pa-$\beta$ line intensity ratios of 
various types of astronomical sources.  
Referenecs: Cas A \cite{koo2013, lee2014}, G11.2-0.3 (Moon, D.-S. personal communication), 
RCW 103 \citep{oliva1989, oliva1990}, LMC SNRs \citep{oliva2001},
HH46-47 \citep{garcia2010}, Orion \citep{walmsley2000}, and NGC 1068 \citep{oliva2001}.
\label{fig:pkasfig4}}
\end{figure}

%%% CALL LIST OF REFERENCES (natbib STYLE) %%%%%%%%%%%%%%%%%%%%%%%%%%

\end{document}